\documentclass[10pt,amssymb,final,JAP]{iopart} 

\usepackage{iopams}  
\usepackage[utf8]{inputenc}
\usepackage{ulem}
\usepackage{graphicx}   
\usepackage{dcolumn}
\usepackage{bm}
\usepackage{setspace}

\usepackage{color}
\begin{document}


\title[Influence of deep levels on the electrical transport properties of CdZnTeSe detectors]{Influence of deep levels on the electrical transport properties of CdZnTeSe detectors}

\author{M Rejhon$^1$, J Franc$^1$, V D\v{e}di\v{c}$^1$, J Pekárek$^{1}$, U N Roy$^2$, R Grill$^1$ and R B James$^3$}
\ead{rejhonm@karlov.mff.cuni.cz}
\address{$^1$ Charles University, Faculty of Mathematics and Physics, Institute of Physics, Ke Karlovu 5, Cz-121 16 Prague 2, Czech Republic}
\address{$^2$ Nonproliferation and National Security Department, Brookhaven National Laboratory, Upton, New York, USA}
\address{$^3$ Savannah River National Laboratory, Aiken, South Carolina 29808, USA}
\date{\today}

\begin{abstract}
We investigated the influence of deep levels on the electrical transport properties of CdZnTeSe radiation detectors by comparing experimental data with numerical simulations based on simultaneous solution of drift-diffusion and Posisson equations, including the Shockley-Read-Hall model of the carrier trapping. We determined the Schottky barrier height and the Fermi level position from I-V measurements.  We measured the time evolution of the electric field and the electrical current after application of a voltage bias. We observed that the electrical properties of CZTS are fundamentally governed by two deep levels close to the mid-bandgap - one recombination and one hole trap. We show that the hole trap indirectly increases the mobility-lifetime product of electrons. We conclude that the structure of deep levels in CZTS are favorable for high electrical charge transport.
\end{abstract}
                             

                            
\submitto{\JAP}


\section{Introduction}
CdZnTeSe (CZTS) has been a focus of research interest as a material for hard X-ray and gamma-ray room temperature semiconductor detectors \cite{BEZSMOLNYY2001,Roy2015,Roy2015APL}. So far, the semiconductor material of choice is CdZnTe (CZT) with 10-15$\%$ of Zn content \cite{SCHLESINGER2001,Iniewski2016,wahl2014,Krawczynski2016}. It was demonstrated that the addition of Se to the  CZT matrix results in effective lattice hardening and a reduction of Te inclusions and dislocations, which did not form a cellular structure like CZT \cite{TANAKA1989,Gul2017}. Apparently, it is a consequence of a decrease in the lattice constant by $0.41$~\AA /mol \cite{TANAKA1989}.

It was also observed that CZTS crystals exhibit better crystallinity than CZT, which can lead to a larger yield of high-quality material with comparable electrical and spectroscopic properties as CdTe and CZT materials \cite{Gul2017}.

The electrical charge transport in semiconductor detectors is affected by the material properties, such as deep levels and metal-semiconductor interfaces. The deep levels can act as trapping centers or recombination centers \cite{BUIS2014188,BOLOTNIKOV201346}. These are formed by the presence of defects, such as sub-grain boundaries, Te inclusions/precipitates, impurities and compositional inhomogeneity \cite{MacKenzie2013,BURGER2000586}.

In the current study we investigated the deep levels that influence the time and temperature dependence of the electrical current, carrier trapping and recombination of photo-generated charge. We applied measurements of I-V characteristics and electro-optic Pockels effect in this work. We also simulated the results using  drift-diffusion and Poisson equations, including the influence of the Shockley-Read-Hall model. 
\section{Experimental}
We investigated a CZTS sample with $4\%$ Se and $10\%$ Zn fabricated by the travelling heat method (THM). We chose a sample with a moderate $\mu\tau$ product for electrons ($\approx1\times10^{-3}$~cm$^{2}$/Vs) in order to understand how deep levels influence its electron-transport properties and may limit the detector performance. The sample dimensions were $6.50\times5.30\times2.68$~mm$^{3}$. The sample surfaces were mechanically polished with Al$_2$O$_3$ abrasive (surface RMS of $2$~nm) without any further chemical treatment. To measure the I-V characteristics and electric field profile, we deposited gold and indium contacts on the larger opposite sides by evaporation. The motivation to use different metals for the opposite contacts was derived from our experience with CZT, where  band bending upwards occurs at the Au/CZT interface and downwards at the In/CZT interface. The Au cathode then blocks electrons while the In anode blocks holes. In this way a decrease of the dark current can be achieved independent of the position of the Fermi level, which is located close to mid-bandgap for this high-resistivity material. The gold contact was equipped with a guard ring to separate the bulk and surface leakage currents during the electrical measurements.

The I-V characteristics were measured at $300$~K. A Keithley 2410 sourcemeter was used to bias the sample, and a Keithley 6514 electrometer was used to measure the bulk current on a serial $1$~M$\Omega$ resistor. The sample resistivity was evaluated from the I-V curve around zero voltage, and the Schottky barrier height was determined from the I-V characteristics at high applied voltages.

\begin{figure}[htbp]
\centering
\includegraphics[width=6cm]{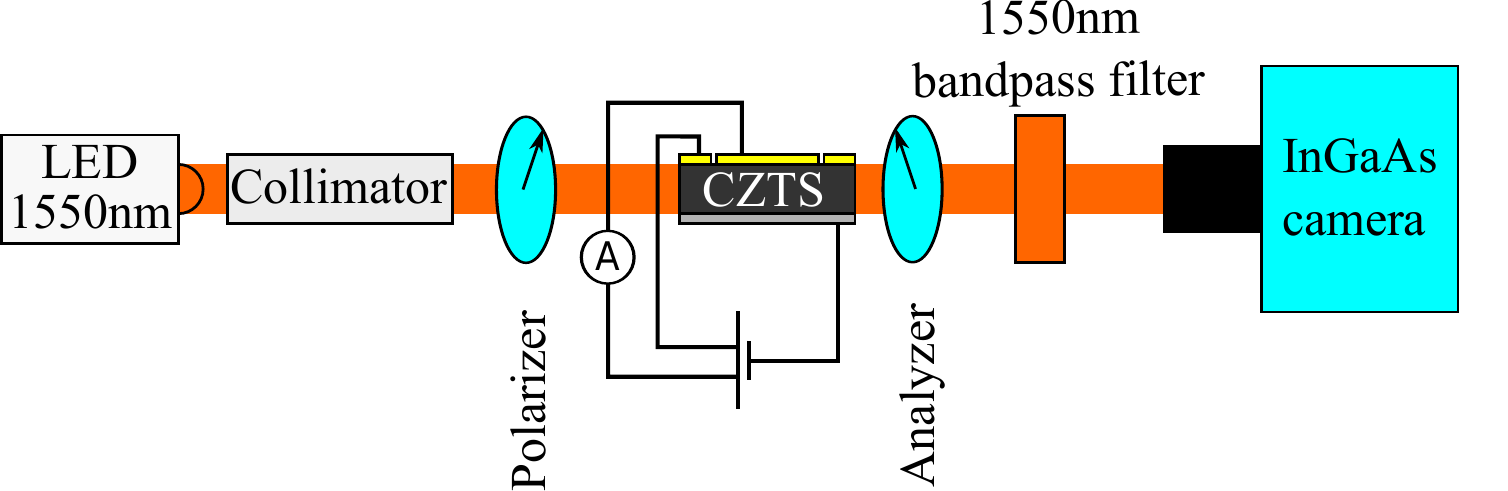}
\caption{Scheme of the experimental setup used for measurement of the electric field in CZTS.}
\label{schem}
\end{figure}

We evaluated the time evolution of the electric field and the electrical current after application of a voltage bias at temperatures of $290$~K, $300$~K and $310$~K. The electro-optic Pockels effect was used to measure the electric field in the CZTS sample. A standard setup consisted of a testing light, two polarizers and an InGaAs camera. The sample was placed between two crossed polarizers, and testing light at a wavelength of $1550$~nm passed through the sample. The transmitted light distribution $T(x,y)$ was collected by an InGaAs camera (figure \ref{schem}). The electric field distribution $\mathcal{E}(x,y)$ depends on the transmittance as $\mathcal{E}(x,y)\sim\arcsin\sqrt{T(x,y)}$. More details about the measurement setup and evaluation of the electric field profiles can be found in our previous articles \cite{rejhonJPD,Franc2015}.

The time and temperature evolutions of the electric field and the current after the application of a voltage bias were compared with a numerical simulation based on the solution of the drift-diffusion and Poisson equations, including the Shockley-Read-Hall model \cite{Grill2012}.

\section{Results and Discussion}
\subsection{I-V measurements}

\begin{figure*}[htbp]
\centering
\includegraphics[width=12cm]{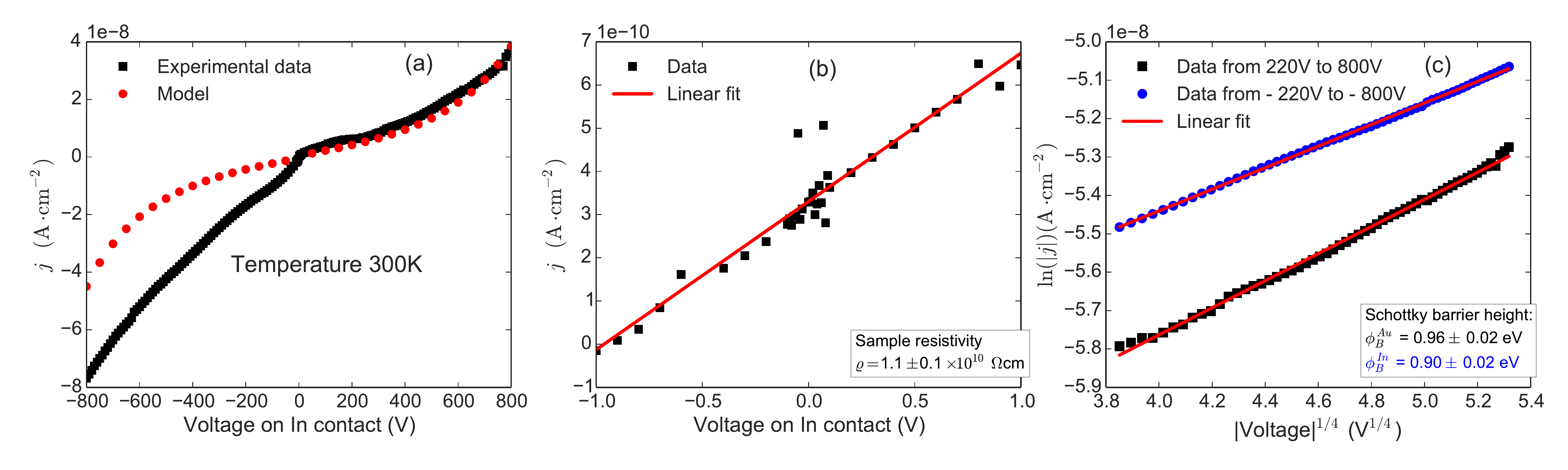}
\caption{The I-V characteristics of In/CZTS/Au sample measured at $300$~K (black squares) and simulated I-V curve (red circles) are plotted in graph (a). The graph (b) shows the I-V curve around zero voltage with a linear fit (Ohm's law). A sample resistivity $\varrho=1.10\times10^{10}$~$\Omega$cm  was determined. The Schottky barrier heights were computed using Eq. (\ref{Eq_bar}) from the experimental data (graph (c)).}
\label{IV}
\end{figure*}

The I-V measurement of bulk current in the range from $-800$~V up to $+800$~V at $300$~K is shown in figure \ref{IV}(a). We evaluated the sample resistivity $\varrho=1.1\times10^{10}$~$\Omega$cm from the I-V curve between $-1$ and $+1$~V using Ohm's law (figure \ref{IV}(b)). The Fermi level $E_{F}$ is connected to resistivity by the equation:

\begin{equation}
\varrho=\frac{1}{\left[\mu_{e}N_{c}\exp{\left(\frac{E_{F}}{kT}\right)}+ \mu_{h}N_{v}\exp{\left(\frac{(-E_{F}-E_{G})}{kT}\right)}\right]},
\label{Eq_fermi}
\end{equation}
where $\mu_{e}$ and $\mu_{h}$ are the electron and hole mobility, respectively. $N_{c,v}$ is the effective density of states for electrons (holes) in the conduction (valence) band, $k$ is the Boltzmann constant, $T$ is the absolute temperature and $E_{G}$ is the bandgap energy ($E_{G}=1.52$~eV \cite{rejhonArxiv}). The general carrier mobilities for electrons $\mu_{e}\approx1100$~cm$^{2}$V$^{-1}$s$^{-1}$ and for holes $\mu_{h}\approx80$~cm$^{2}$V$^{-1}$s$^{-1}$ were used \cite{Pousset2016,Pavesi2017}. We observed from thermoelectric power measurements a slight n-type conductivity for sample. Therefore, we assume that the calculated Fermi level is related to the conduction band minimum at $E_{F}=E_{c}-0.727$~eV.

The I-V curve at high voltage due to the Schottky barrier height $\phi_{B}$ and Schottky barrier lowering $\Delta\phi_{B}$ is given by the equation \cite{Sze2007}
\begin{equation}
j=A^{\star}T^{2}\exp\left(-\frac{\phi_{B}-\Delta\phi_{B}}{kT}\right),
\label{Eq_bar}
\end{equation}
where $A^{\star}$ is the Richardson constant. The Schottky barrier lowering in the simplest form is $\Delta\phi_{B}=e\sqrt{\frac{eE}{4\pi\varepsilon}}$ and the electric field is $E=\sqrt{\frac{2eN_{t}}{\varepsilon}U}$. Here, $e$ is the elementary charge, $\varepsilon$ is the absolute permittivity, $N_{t}$ is the deep level concentration, and $U$ is the applied bias. If the current dependence $\ln(j)$ versus $U^{1/4}$ is linear, this model based on Schottky barrier height and barrier lowering due to the applied high voltage is valid.

The relation between $\ln(j)$ and $U^{1/4}$ for voltages in the range of $220-800$~V for both polarities is presented in figure \ref{IV}(c).

The current flow at high voltage is limited by the Schottky barrier height at both polarities. It means that the bands bend at both interfaces (Au/CZTS and In/CZTS) upwards. We evaluated the Schottky barrier height at the Au/CZTS interface as $\phi_{B}^{Au}=0.96\pm0.02$~eV and at the In/CZTS interface as $\phi_{B}^{In}=0.90\pm0.02$~eV. 

\subsection{Dynamics of the electric field and current plus simulations}
\begin{figure}[htbp]
\centering
\includegraphics[width=6cm]{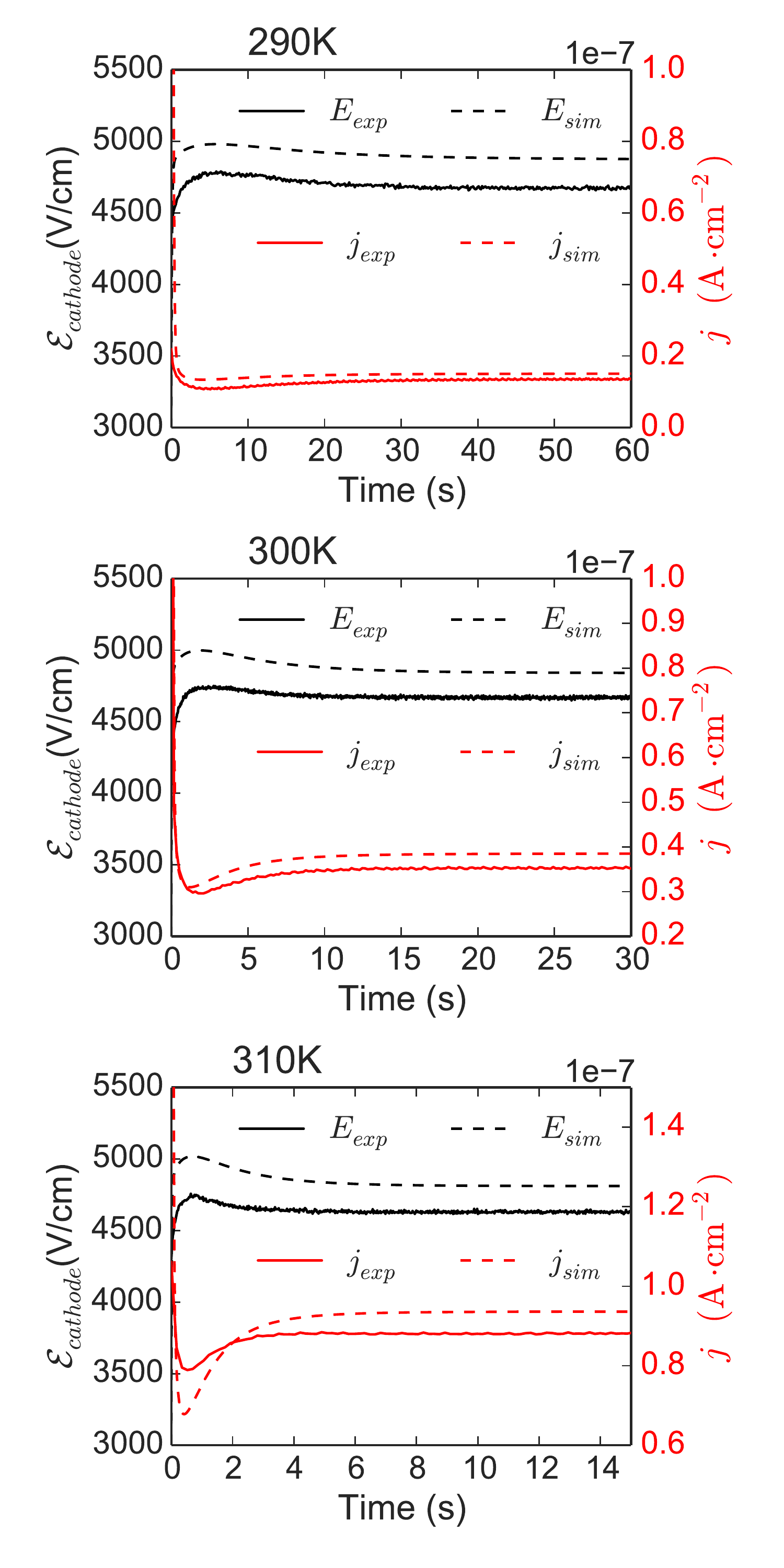}
\caption{Time evolution of the electric field under the cathode (black solid line) and the electric current (red solid line) after switching on the bias at different temperatures. The numerical simulation of the electric field evolution under the cathode is represented by a black dashed line and the current is represented by red dashed line.}
\label{el}
\end{figure}

We also measured the time evolution of the electrical current after application of a bias, together with measurements of the internal electric field at $290$~K, $300$~K and $310$~K using the electro-optic Pockels effect. The measured data of the bulk electrical current and electric field under the cathode are shown in figure \ref{el} (solid lines).

The electrical current strongly decreases within approximately the first two seconds. Then, it starts to increase over time. The electric field under the cathode has an opposite course (it inceases within the first two seconds, and then it starts to decrease). 

We simulated the temporal evolution of the electric field and the electrical current using the solution of the drift-diffusion and Poisson equations including the Shockley-Read-Hall model \cite{Hall52,SR52}. The simulated data are depicted in figure \ref{el} (dashed lines). The parameters used for the simulation are listed in Table \ref{tab1}. We were able to fix a number of parameters from experimental data, and in this way we increased the reliability of the simulations. The Schottky barrier heights at the In/CZTS and Au/CZTS interfaces were set from the evaluation of the I-V characteristics (figure \ref{IV}). The band-gap energy was set from the results of ellipsometry measurements \cite{rejhonArxiv}. The Fermi level position was calculated according to Eq. (\ref{Eq_fermi}). The energies of the deep levels were set based on results of the evaluation of temperature and time evolution of the electrical field by Pockels effect studied in detail in our article \cite{rejhonArxiv}. The capture cross-sections of deep levels were used as fitting parameters. We applied the experimentally determined values of electron capture cross-section of level $E_{1}$ and hole capture cross-section of level $E_{2}$ as initial values for fitting. The resulting optimal values (Table \ref{tab1}) are $1-2$ orders of magnitude smaller. These set of fitting parameters presented in Table \ref{tab1} lead to a very good agreement between experiment and simulated data (figure \ref{el}). According to simulation results the level $E_{1}$ is acting as a recombination center with a higher capture cross-section for electrons than for holes. The deep level $E_{2}$ is in the model a pure hole trap.

\begin{table*}
\caption{Parameters of the deep levels using in the simulations}
\begin{tabular}{lc||lc|c}
\hline
\hline
General condition: 		& 							& Deep levels:	&	$E_{1}$	& $E_{2}$	\\
Bias on In electrode (V)	& $800$  						& Energy (eV) 		& 	$E_c-0.836$	& $E_c-0.697$	\\
Sample properties: 		& 							& Concentration (cm$^{-3}$)		& 	$2.8 \times10^{11}$			&	$5.4 \times10^{11}$	\\
Thickness (mm)				&  		$2.68$					& $\sigma_{n}$	(cm$^{2}$)	& $5.0\times10^{-13}$				&	$1.1\times10^{-25}$	\\
Bandgap energy (eV)	& $1.52$ 					& $\sigma_{p}$ 	(cm$^{2}$)	& 		$7.5\times10^{-16}$	 		& $7.0\times10^{-15}$		\\
Fermi level energy (eV)		& $E_{F}=E_{c}-0.727$	& 		& 				&	\\
Schottky barrier height (eV)		& $\phi_{Au}=0.96$/$\phi_{In}=0.90$	& 		& 				&	\\
\hline
\hline

\end{tabular}
\label{tab1}
\end{table*}

\begin{figure*}[htbp]
\centering
\includegraphics[width=12cm]{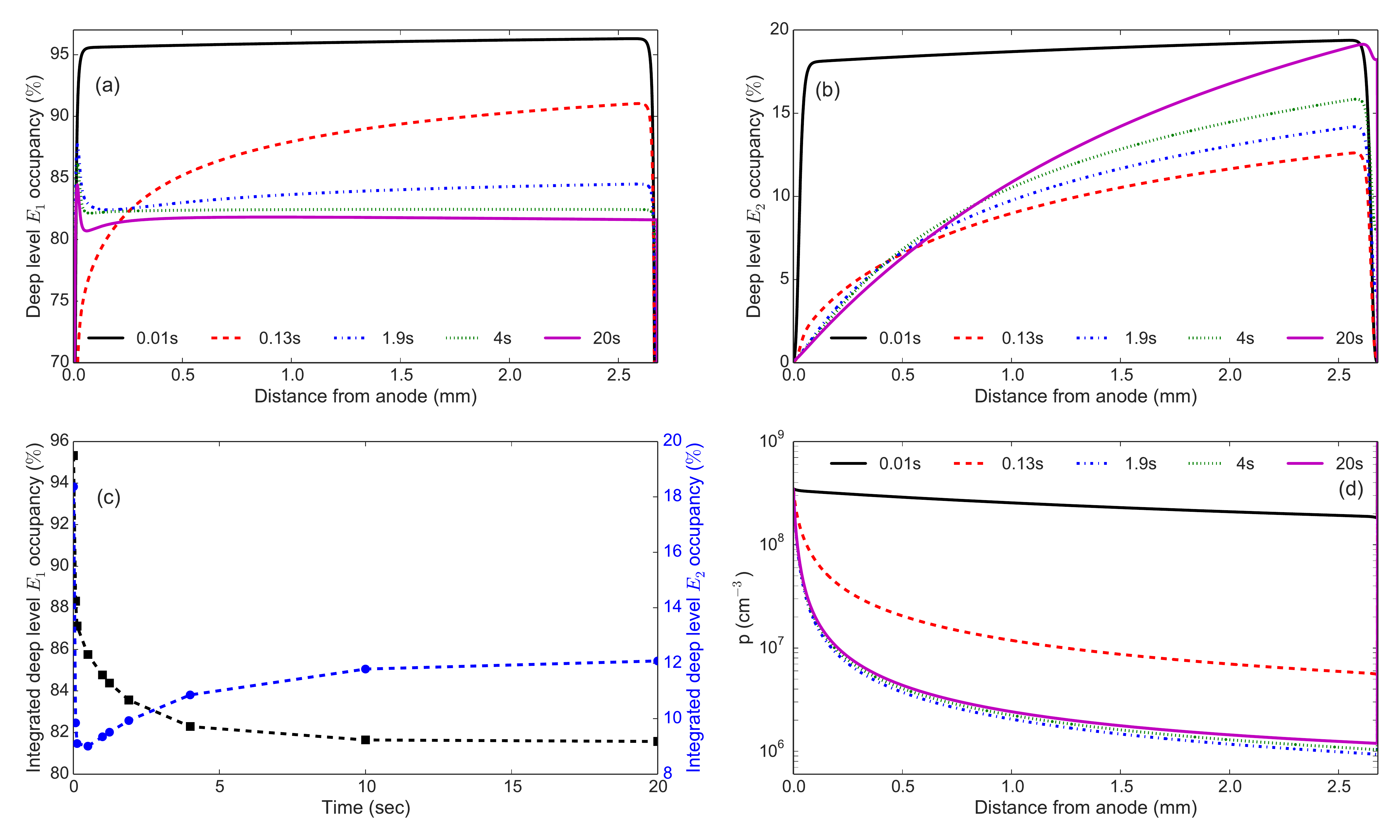}
\caption{The graph (a) shows the deep level $E_{1}$ occupancy. The deep level $E_{2}$ occupancy is depicted on graph (b). The graph (c) exhibits the integrated deep level occupancy for deep level $E_{1}$ as black squares and for deep level $E_{2}$ as blue circles. The distribution of free holes in the valence band is plotted on graph (d).}
\label{occup}
\end{figure*}

The results of the simulation show that the electrical current is predominantly formed by the flow of injected free holes due to the band bending upwards at the interfaces. The Au/CZTS interface blocks electrons in the regime of application of a positive voltage on the In contact. The In/CZTS interface injects the holes into the volume of the sample. The damped oscillatory character with undershoots of both the electric current and of the electric field under the cathode is caused by the change of occupancy of deep levels over time. The computed deep level's occupancies by electrons  are depicted in figure \ref{occup}. The presented profiles and their evolution in time can be described by the following series of processes.

\begin{figure}[htbp]
\centering
\includegraphics[width=6cm]{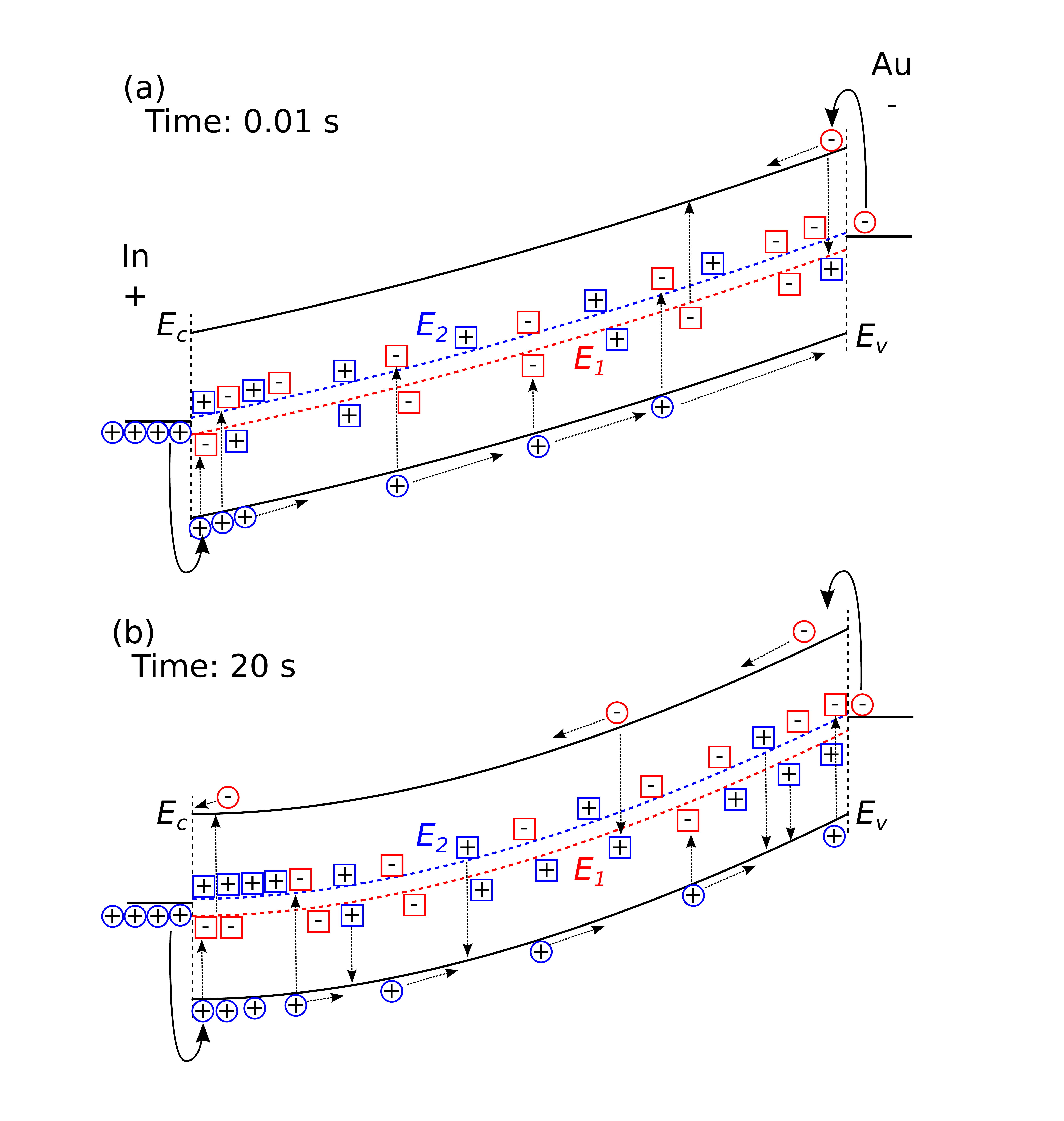}
\caption{Simplified scheme of trapping and de-trapping of carriers. Scheme (a) is plotted for time $0.01$~s after bias application. The carriers are injected from the contact into the bulk, where they are trapped at the deep levels. Scheme (b) shows the steady- state at $20$~s after switching on the bias.}
\label{schema}
\end{figure}

At first holes are injected from the In anode within $0.1$~sec after application of the bias and are captured on both deep levels. This effect causes a strong decrease in the occupation of both deep levels, especially for those close to the anode (see the dashed lines showing the occupancy profiles after $0.13$~sec in figures \ref{occup}(a,b)). The simplified scheme describing these processes is plotted in figure \ref{schema}(a). This leads to an increase of the positive space charge in the sample.  Due to the formed positive space charge,  the electric field under the cathode increases. After that, the higher electric field depletes the holes from the volume near the cathode, which leads to a higher emission of captured holes from the deep level $E_2$  accompanied by an increase of its occupancy (see profiles after $1.9$~sec and $4$~sec in figure \ref{occup}(b) and schemes in figure \ref{schema}(b)). This process causes the subsequent decrease of the positive space charge, which induces a decrease of the electric field below the cathode. The electrical current increases due to the higher amount of free holes. 

The graph \ref{occup}(c) shows that the deep level $E_1$ is monotonously discharged (electron transfer from the level) over time (black squares with dashed line). On the other hand, the integrated occupancy of the deep level $E_2$ exhibits an undershoot, which induces the undershoot in the electric field. The profiles of the free holes in the valence band are depicted in figure \ref{occup}(d).

The I-V curve was simulated by using the solution of the drift-diffusion with respect to a Schottky barrier lowering and Poisson equations including the Shockley-Read-Hall model. The result of the simulated I-V curve is depicted in figure \ref{IV}(a). There is a good agreement in the regime with a positive voltage on the In contact. The opposite regime (negative voltage on the In contact) shows a mismatch between experiment and simulation. It can be caused by a more complex character of the defect structure of the material extending beyond the simplified two-level model. However, we assume that these two deep levels are the main ones responsible for the undershoot in the electric field and of the current evolution.

\begin{figure}[htbp]
\centering
\includegraphics[width=6cm]{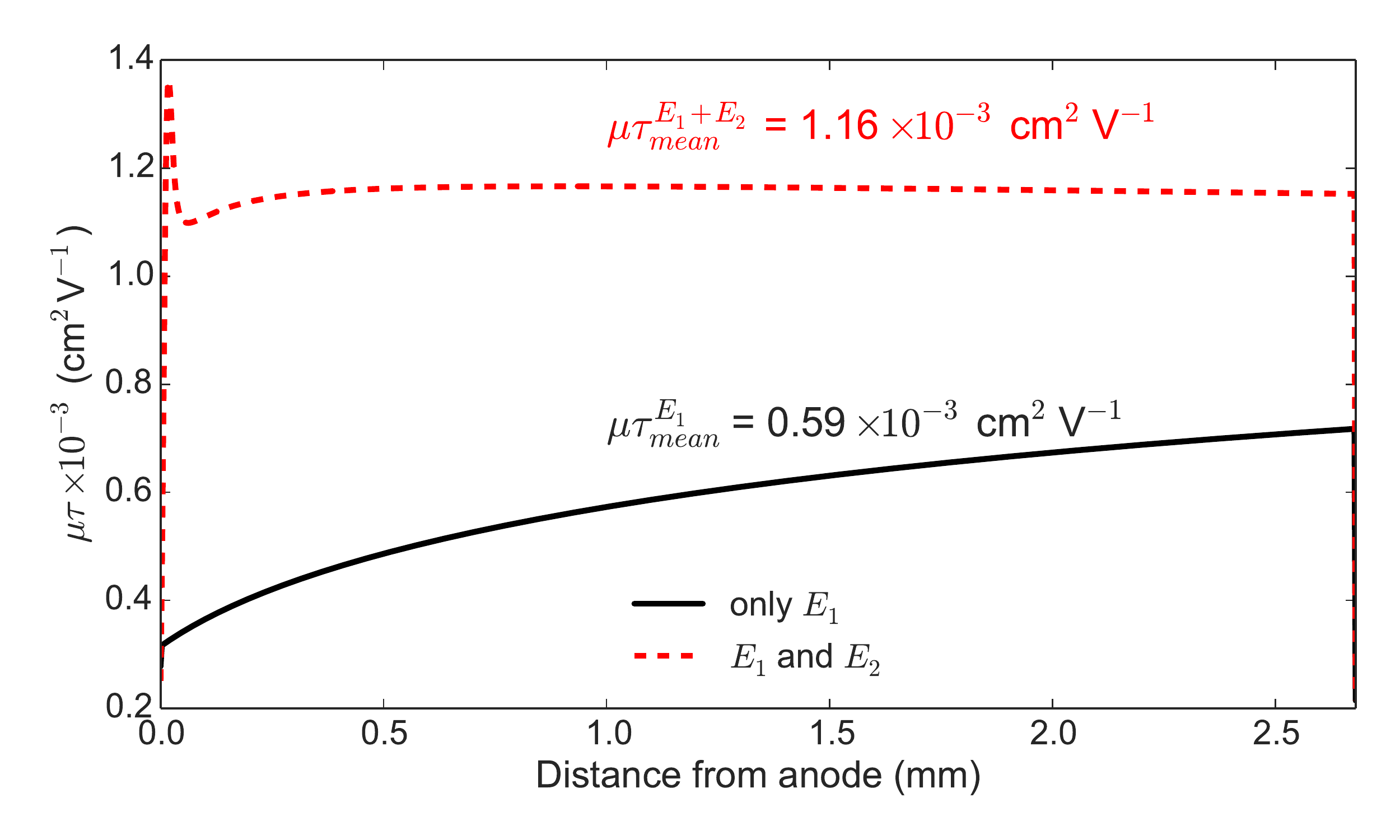}
\caption{The profiles of the electron $\mu\tau$ product based on simulations with and without the deep level $E_{2}$.}
\label{mutau}
\end{figure}

Figure \ref{mutau} shows the results of simulations for the electron $\mu\tau$ product for the case of only one level $E_{1}$ and the case with both levels $E_{1}+E_{2}$. The mean of the electron $\mu\tau^{E_{1}}$ product is $0.59\times10^{-3}$~cm$^{2}$V$^{-1}$. The electron $\mu\tau^{E_{1}+E_{2}}$ product for the case with both levels is $1.16\times10^{-3}$~cm$^{2}$V$^{-1}$, which agrees with the experimentally determined value $\mu\tau^{exp}_{e}=1.10\times10^{-3}$~cm$^{2}$V$^{-1}$. The electron $\mu\tau$ product is higher for the case with both levels due to the presence of the hole deep level $E_{2}$. This level communicates dominantly with the valence band and acts as a hole trap. It reduces the concentration of free holes and in this way also transitions electrons from the recombination trap $E_{1}$ to the valence band. The occupancy of the level $E_{1}$ with electrons is increased, transitions of electrons from the conduction band to the level $E_{1}$ decreased and the mobility-lifetime product of electrons increased.

\section{Conclusions}

We investigated the deep levels that influence the carrier transport and performance of detector-grade CZST using the I-V characteristics, Pockels effect and numerical simulations based on the Shockley-Read-Hall model. We determined  the Schottky barrier height and the Fermi level position in the CZTS sample. These parameters describing the sample, supplemented with our previous results about deep levels in the material, were used in the simulation of the electric field and electrical current evolution over time. We achieved a very good agreement between the experimental and simulated data using a model with one deep recombination and one deep hole trap. The hole trap increases the mobility-lifetime product of electrons by decreasing the concentration of free holes and in this way increasing the electron occupancy of the deep recombination trap. We conclude that the deep levels in CZTS have a favourable structure for high electric-charge transport, which is critical for producing high-performance X-ray and gamma-ray detectors.

\section*{Acknowledgments}
This work was supported by the U.S. Department of Energy, Office of Defense Nuclear Nonproliferation Research and Development, DNN R$\&$D. This paper was also financially supported by the  Grant Agency of Czech Republic (GAČR), project 102-18-06818S and student project SVV–2018–260445.

\bibliographystyle{iopart-num}
\section*{References}
\providecommand{\newblock}{}


\begin{thebibliography}{10}
\expandafter\ifx\csname url\endcsname\relax
  \def\url#1{{\tt #1}}\fi
\expandafter\ifx\csname urlprefix\endcsname\relax\def\urlprefix{URL }\fi
\providecommand{\eprint}[2][]{\url{#2}}

\bibitem{BEZSMOLNYY2001}
Bezsmolnyy Y 2001 {\em Nuclear Instruments and Methods in Physics Research
  Section A: Accelerators, Spectrometers, Detectors and Associated Equipment\/}
  {\bf 458} 461 -- 463 ISSN 0168-9002 proc. of the 11th Internat. Workshop on
  Room Temperature Semiconductor X- and Gamma-Ray Detectors and Associated
  Electronics

\bibitem{Roy2015}
Roy U~N, Bolotnikov~A E, Camarda G, Cui Y, Hossain A, Lee K, Lee W, Tappero R,
  Yang G, Cui Y, Burger A and James R~B 2015 {\em Journal of Crystal Growth\/}
  {\bf 411} 34--37 ISSN 00220248

\bibitem{Roy2015APL}
Roy U~N, Bolotnikov A~E, Camarda G~S, Cui Y, Hossain A, Lee K, Lee W, Tappero
  R, Yang G, Gul R and James R~B 2015 {\em APL Materials\/} {\bf 3} 026102
  (\textit{Preprint} \eprint{https://doi.org/10.1063/1.4907250})

\bibitem{SCHLESINGER2001}
Schlesinger~T E, Toney~J E, Yoon H, Lee E, Brunett B, Franks L and James R~B
  2001 {\em Materials Science and Engineering: R: Reports\/} {\bf 32} 103 --
  189 ISSN 0927-796X

\bibitem{Iniewski2016}
Iniewski K 2016 {\em Journal of Instrumentation\/} {\bf 11} C12034

\bibitem{wahl2014}
Wahl C~G, Kaye W, Wang W, Zhang F, Jaworski J, Boucher Y~A, King A and He Z
  2014 Polaris-h measurements and performance {\em 2014 IEEE NUCLEAR SCIENCE
  SYMPOSIUM AND MEDICAL IMAGING CONFERENCE (NSS/MIC)\/} IEEE (345 E 47TH ST,
  NEW YORK, NY 10017 USA: IEEE) ISBN 978-1-4799-6097-2 iEEE Nuclear Science
  Symposium / Medical Imaging Conference (NSS/MIC), Seattle, WA, NOV 08-15,
  2014

\bibitem{Krawczynski2016}
Krawczynski H~S, Stern D, Harrison F~A, Kislat F~F, Zajczyk A, Beilicke M,
  Hoormann J, Guo Q, Endsley R, Ingram A~R, Miyasaka H, Madsen K~K, Aaron K~M,
  Amini R, Baring M~G, Beheshtipour B, Bodaghee A, Booth J, Borden C, Böttcher
  M, Christensen F~E, Coppi P~S, Cowsik R, Davis S, Dexter J, Done C, Dominguez
  L~A, Ellison D, English R~J, Fabian A~C, Falcone A, Favretto J~A, Fernández
  R, Giommi P, Grefenstette B~W, Kara E, Lee C~H, Lyutikov M, Maccarone T,
  Matsumoto H, McKinney J, Mihara T, Miller J~M, Narayan R, Natalucci L, Özel
  F, Pivovaroff M~J, Pravdo S, Psaltis D, Okajima T, Toma K and Zhang W~W 2016
  {\em Astroparticle Physics\/} {\bf 75} 8--28 ISSN 09276505

\bibitem{TANAKA1989}
Tanaka A, Masa Y, Seto S and Kawasaki T 1989 {\em Journal of Crystal Growth\/}
  {\bf 94} 166 -- 170 ISSN 0022-0248

\bibitem{Gul2017}
Gul R, Roy U~N, Camarda G~S, Hossain A, Yang G, Vanier P, Lordi V, Varley J and
  James R~B 2017 {\em Journal of Applied Physics\/} {\bf 121} 125705
  (\textit{Preprint} \eprint{https://doi.org/10.1063/1.4979012})

\bibitem{BUIS2014188}
Buis C, Aillon E~G, Lohstroh A, Marrakchi G, Jeynes C and Verger L 2014 {\em
  Nuclear Instruments and Methods in Physics Research Section A: Accelerators,
  Spectrometers, Detectors and Associated Equipment\/} {\bf 735} 188 -- 192
  ISSN 0168-9002

\bibitem{BOLOTNIKOV201346}
Bolotnikov A, Camarda G, Cui Y, Yang G, Hossain A, Kim K and James R 2013 {\em
  Journal of Crystal Growth\/} {\bf 379} 46 -- 56 ISSN 0022-0248 compound
  Semiconductors and Scintillators for Radiation Detection Applications: A
  Special Tribute to the Research of Michael Schieber

\bibitem{MacKenzie2013}
MacKenzie J, Kumar F~J and Chen H 2013 {\em Journal of Electronic Materials\/}
  {\bf 42} 3129--3132 ISSN 1543-186X

\bibitem{BURGER2000586}
Burger A, Chattopadhyay K, Chen H, Ma X, Ndap J~O, Schieber M, Schlesinger T,
  Yao H, Erickson J and James R 2000 {\em Nuclear Instruments and Methods in
  Physics Research Section A: Accelerators, Spectrometers, Detectors and
  Associated Equipment\/} {\bf 448} 586 -- 590 ISSN 0168-9002

\bibitem{rejhonJPD}
Rejhon M, Franc J, Dědič V, Kunc J and Grill R 2016 {\em Journal of Physics
  D: Applied Physics\/} {\bf 49} 375101

\bibitem{Franc2015}
Franc J, Dědič V, Rejhon M, Zázvorka J, Praus P, Touš J and Sellin P~J 2015
  {\em Journal of Applied Physics\/} {\bf 117} 165702

\bibitem{Grill2012}
Grill R, Franc J, Elhadidy H, Belas E, \u{S} Uxa, Bugar M, Moravec P and Hoschl
  P 2012 {\em IEEE Transactions on Nuclear Science\/} {\bf 59} 2383--2391 ISSN
  0018-9499

\bibitem{rejhonArxiv}
Rejhon M, Dědič V, Beran L, Roy U~N, Franc J and James R~B 2018 Investigation
  of deep levels in cdzntese crystal and their effect on the internal electric
  field of cdzntese gamma-ray detector (\textit{Preprint}
  \eprint{arXiv:1809.10353})

\bibitem{Pousset2016}
Pousset J, Farella I, Gambino S and Cola A 2016 {\em Journal of Applied
  Physics\/} {\bf 119} 105701

\bibitem{Pavesi2017}
Pavesi M, Santi A, Bettelli M, Zappettini A and Zanichelli M 2017 {\em IEEE
  Transactions on Nuclear Science\/} {\bf 64} 2706--2712 ISSN 0018-9499

\bibitem{Sze2007}
Sze S and Ng K~K c2007 {\em Physics of semiconductor devices\/} 3rd ed
  (Hoboken, N.J.: Wiley-Interscience) ISBN 978-0-471-14323-9

\bibitem{Hall52}
Hall R~N 1952 {\em Phys. Rev.\/} {\bf 87}(2) 387--387

\bibitem{SR52}
Shockley W and Read W~T 1952 {\em Phys. Rev.\/} {\bf 87}(5) 835--842

\end{thebibliography}
\end{document}